\documentclass[prl,twocolumn,showpacs,groupedaddress]{revtex4}
\usepackage[dvips]{graphicx}
\usepackage{graphicx}

\begin{document}
\title{Analysis of photon-atom entanglement generated by
Faraday rotation in a cavity}
\author{S. K. Y. Lee and C. K. Law}
\affiliation{Department of Physics and Institute of Theoretical
Physics, The Chinese University of Hong Kong, Shatin, Hong Kong SAR,
China}
\date{\today}
\begin{abstract}
Faraday rotation based on AC Stark shifts is a mechanism that can
entangle the polarization states of photons and atoms. We study the
entanglement dynamics inside an optical cavity, and characterize the
photon-atom entanglement by using the Schmidt decomposition method.
The time-dependence of entanglement entropy and the effective
Schmidt number are examined. We show that the entanglement can be
enhanced by the cavity, and the entanglement entropy can be
controlled by the initial fluctuations of atoms and photons.

\end{abstract}
 \maketitle

The interaction between quantized optical fields and atomic
ensembles provides physical models for exploring and realizing novel
applications in quantum communication. A useful mechanism is the AC
Stark shifts of off-resonant light with two polarizations. As the
frequency shift in each polarization is conditioned by the number of
atoms at the corresponding magnetic sub-levels, any imbalance of
atom numbers would lead to a rotation of Stokes parameters of light.
The effect can be considered as a kind of Faraday rotation that
correlates the polarization variables of photons and atoms, and it
has been investigated theoretically and experimentally in the
context of QND measurements and spin squeezing
\cite{sqz1,sqz2,Oblak}. Interesting applications based on the
interaction have also been discussed in recent literature. These
include, for examples, quantum states teleportation and swapping
\cite{teleport}, quantum memory \cite{memory}, generation of
macroscopic superposition states \cite{macroscopic}, and the
implementation of Deutsch-Jozsa algorithm \cite{agarwal}.

The key to most of these intriguing applications is the
establishment of quantum entanglement between photons and atoms,
governed by a model Hamiltonian \cite{sqz1}:
\begin{equation}\label{1}
H_I =2 g S_z J_z,
\end{equation}
where $S_z$ is one of the Stokes operators of the field, $J_z$ is a
collective spin operator of atoms, and $g$ is a coupling constant.
Such an interspecies entanglement can be further converted into
non-classical spin states by projective measurement of the light
field. Previous studies of the interaction (1) have emphasized the
quantum correlations of observables relevant to QND experiments
\cite{sqz1,sqz2,Oblak}. From the view point of quantum information,
a better understanding of entanglement can be gained by analyzing
directly the photon-atom state vector in the Hilbert space. In
particular, it would be useful to quantify the degree of
entanglement and determine what parameters controlling the
entanglement dynamics.

In this paper we approach the problem by studying the Schmidt
decomposition of entangled states generated by (1). The distribution
of Schmidt eigenvalues and the structure of Schmidt eigen-modes
provide a complete characterization of pure-state entanglement
\cite{knight}. In contrast to usual investigations in free space, we
will examine the interaction (1) in an optical cavity. This is
motivated by the fact that a strong atom-field coupling is often
desirable to overcome decoherence. In free space, although the
coupling strength can be increased at frequencies near resonances,
the higher excitation rate would lead to substantial decoherence
from spontaneous emission \cite{Oblak,hammerer}. The employment of a
cavity provides a way to enhance the coupling strength while keeping
the fields far away from resonance. In addition, the strong coupling
may also allow us to explore cavity QED effects with fewer number of
photons and atoms.

\begin{figure}
\includegraphics[width=6 cm]{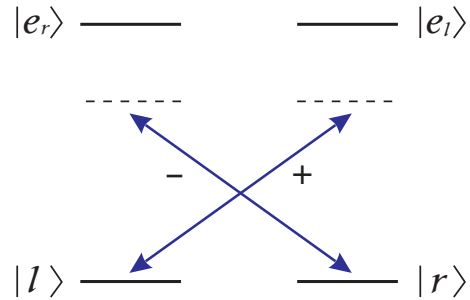}
\caption{(Color online) Atomic levels and interaction scheme of our
model Hamiltonian. AC Stark shift is induced by non-resonant
coupling between ground state $| l \rangle$ ($| r \rangle$) and
polarization mode $+$ ($-$). For example, $| l \rangle$ and $| r
\rangle$ refer to the two ground states with magnetic quantum
numbers $\pm 1/2$.} \label{fig1}
\end{figure}

Our system consists of $N_A$ four-level atoms interacting with a
quantized cavity field with two polarizations. The two degenerate
atomic ground states are denoted by $| l \rangle$ and $| r
\rangle$, and they are coupled to the excited states $| e_l
\rangle$ and $| e_r \rangle$ by absorbing or emitting a photon of
the corresponding polarization (Fig. 1). In our scheme, the field
frequency is significantly detuned from the atomic transition
frequency, so that the excited states are rarely populated. Hence
the decoherence effects due to spontaneous decay from $| e_l
\rangle$ and $| e_r \rangle$ can be safely neglected. The
off-resonant interaction leads to AC Stark shifts, which are
described by the effective Hamiltonian \cite{agarwal}:
\begin{equation}
H = g \sum\limits_{k=1}^{N_A}  \left(  a_+^{\dag} a_+ \left| l
\right\rangle _{kk} \left\langle l \right|
 +  a_-^{\dag} a_-
\left| r \right\rangle _{kk} \left\langle r \right| \right)
\end{equation}
with $ a_+$ and $a_-$ being the annihilation operators of the $+$
and $-$ polarizations of the cavity field. The coupling constant
$g=-|\Omega_0|^2/\Delta$ is obtained by adiabatic elimination of the
excited levels in the large detuning limit, $\Delta \gg |\Omega_0|$,
where $|\Omega_0|$ is the vacuum Rabi frequency defined by the
cavity mode. In our model, the atom-field coupling strength $g$ is
assumed to be the same for each atoms. A ring cavity supporting
traveling wave modes is more suitable to realize our model. This is
because $g$ depends on the absolute square of mode amplitude, a
traveling wave mode would give roughly the same value of $g$ inside
the ring cavity.

By letting $\sigma _z^{(k)}=| l \rangle _{kk} \langle l |- | r
\rangle _{kk} \langle r |$, $J_z=\frac{1}{2}\sum_k {\sigma _z^{(k)}
}$ and $S_z=\frac{1}{2}(a_+^{^\dag}  a_+ - a_-^{^\dag} a_- )$, the
Hamiltonian (2) can be expressed as: $H = \frac{g}{2} N_A
(a_+^{\dag} a_+ + a_-^{\dag}a_- )  + 2 g S_z J_z$. Since the first
term corresponds to a constant shift of field frequency for a fixed
particle number, it can be eliminated in a rotating frame. Faraday
rotation is described by the second term, which is the same as (1).
Noticing that $J_z$ and $S_z$ are formally equivalent to angular
momentum operators, the corresponding eigenvectors are defined by:
$J_z | {j ,m} \rangle =m| {j ,m} \rangle$, $J^2 | {j ,m} \rangle
=j(j+1)| {j ,m} \rangle$, $S_z | { \frac{s}{2} ,\frac{n}{2}} \rangle
=\frac{n}{2} | \frac{s}{2},\frac{n}{2} \rangle$, $S^2 |
{\frac{s}{2},\frac{n}{2}} \rangle =\frac{s}{2}(\frac{s}{2}+1) |
\frac{s}{2} ,\frac{n}{2} \rangle$. Here $2j=N_A$ is the number of
atoms, which is fixed in our problem. The $n$ and $s$ are difference
and sum of photon numbers in the two polarization modes
respectively.

Initially, the system is in a product (disentangled) state:
\begin{equation}\label{intfunc}
\left| {\Psi (0)} \right\rangle  = \left(\sum\limits_{m=-j}^j A _m
\left| {j,m} \right\rangle \right) \otimes \left(
\sum\limits_{s=0}^{\infty} {\sum\limits_{n=-s}^{s}}' P_{s,n}
\left| {\frac{s}{2} ,\frac{n}{2}} \right\rangle \right),
\end{equation}
where $A_m$ and $P_{s,n}$ are amplitudes describing the initial
state of atoms and photons respectively, and the primed summation
sign refers to the step size equal to 2, i.e., $
n=-s,-s+2,-s+4,...,+s$. Based on the Hamiltonian (1), the state
vector at time $t$ is given by,
\begin{equation}\label{timfunc}
\left| {\Psi (t)} \right\rangle  = \sum\limits_{s=0}^\infty
{\sum\limits_{n=-s}^{s}}' \sum\limits_{m=-j}^j  {A_m } P_{s ,n}
e^{ - i g  m n t} \left| j,m \right\rangle \left| {\frac{s}{2}
,\frac{n}{2}} \right\rangle,
\end{equation}
where the phase factor $e^{ - i   g m n t}$ is responsible for the
entanglement. Since the quantum number $s$ plays no role in the
phase factor, it is convenient to define a field state $|n
\rangle_f $ by summing $s$ at a given $n$,
\begin{equation}\label{eq5}
| {n } \rangle_f \equiv F _n^{-1} {\sum\limits_{s = \left| n
\right|}^\infty}' { P_{s,n} \left| {\frac{s}{2},\frac{n}{2}}
\right\rangle } \ \ \ \ \ \ \ n=0,1,2,...
\end{equation}
Here $F _n ^{2} = {\sum\limits_{s = \left| n \right|}^\infty}'
{\left| {P_{s,n} } \right|} ^2$ is a normalization factor. With
some careful counting of states, we note that the $s$ summation in
Eq. (\ref{eq5}) refers to the step size equal to 2 (i.e., $s=|n|,
|n|+2,...$). We also note that the orthogonality relation $_f
\langle {n'}| n \rangle _f = \delta _{n',n}$ is satisfied.

The time-dependent state vector (4) can now be expressed in a
compact form:
\begin{equation}\label{intfunc2}
\left| {\Psi (t)} \right\rangle  = \sum\limits_{n=-\infty}^{\infty}
\sum\limits_{m=-j}^j  {A_m } F_n  e^{ - i  g  m n t}  |m \rangle_a |
{n } \rangle_f,
\end{equation}
where $|m \rangle_a$ is a short notation for the atomic state
$|j=N_A/2,m \rangle$. The evolution of the system is periodic with
the period $t=2\pi/g$. However, we shall see below that a
significant degree of quantum entanglement can be established in
shorter time scale, which depends on the fluctuations of atom
(photon) numbers in the each atomic (polarization) states.

To reveal the pairing structure of photon-atom entanglement in the
state (6), it is customary to perform the Schmidt decomposition so
that,
\begin{equation}\label{Schmidtstate}
\left| {\Psi (t)} \right\rangle =\sum\limits_{k=0}^{\infty} \sqrt
\lambda_k |u_k \rangle _a |v_k \rangle _f.
\end{equation}
Here $\lambda _k$ are Schmidt eigenvalues, and $|u_k \rangle _a$
and $|v_k \rangle _f$ are Schmidt eigenmodes of atoms and photons
respectively. Since the Schmidt modes are orthogonal, i.e., $ _a
\langle u_{k'} |u_k \rangle _a = _f\langle v_{k'} |v_k \rangle _f
= \delta_{kk'}$, if the field is found in the state $|v_k
\rangle$, then with certainty the atomic states must be in $|u_k
\rangle $. In addition, the distribution of  Schmidt eigenvalues
provides a measure of the degree of entanglement. This can be
quantified by the entanglement entropy ${\cal S} =-\sum_k
\lambda_k \ln \lambda_k $, but a more convenient measure is the
effective Schmidt number ${\cal K}$ defined by \cite{grobe}
\begin{equation}
{\cal K} = \left( \sum_k \lambda_k^2  \right)^{-1}.
\end{equation}
${\cal K}$ can be interpreted as the `average' number Schmidt
modes in the expansion. The higher the value of ${\cal K}$, the
higher the entanglement, which shares similar features as the
entropy.

The Schmidt decomposition can be carried out by diagonalizing the
reduced density matrix of individual subsystems \cite{knight}. For
general state vectors, however, one needs to perform the
diagonalization numerically. In order to gain analytic insight
about the properties of entanglement, we consider a class of
states in which $A_m$ and $F_n$ are defined by gaussian functions,
and in addition, these amplitudes change smoothly with $m$ and
$n$:
\begin{eqnarray}
&& A_m =   \chi_A e^{-(m-m_0)^2 / \sigma_A^2}\label{Am}
\\ && F_n = \chi_F e^{-(n-n_0)^2 / \sigma_F^2}.
\end{eqnarray}
Here $\chi_A$ and $\chi_F$ are normalization constants, and $m_0$
($n_0$) and $\sigma_A$ ($\sigma_F$) are parameters corresponding
to the peak and width of atomic (photonic) amplitudes. In order to
have $A_m$ described by a whole Gaussian in the range $-N_A/2\le m
\le N_A/2$, we require $N_A /2>2\sigma_A + \left|m_0\right|$ so that
$A_{\pm N_A/2}$ is negligible. We point out that the gaussian
amplitudes (9) and (10) can be employed to capture or approximate
a variety of states. These include atomic spin coherent states
\cite{spincoherent} and coherent states of photons
\cite{coherent}.

Now we make use of an identity derived from Mehler's formula,
\begin{widetext}
\begin{equation}\label{Schdecomp}
\sqrt {\frac{2}{{\pi \sigma _A \sigma _F }}} e^{-x^2 /
\sigma_A^2-y^2 / \sigma_F^2} e^{ - i  g x y t} = \sqrt {1 -
\mu^2(t)}  \sum\limits_{k = 0}^\infty \mu^{k}(t) U_k(x,t)V_k(y,t)
\end{equation}
\end{widetext}
where $U_k(x,t)$ and $V_k(y,t)$ are oscillator mode functions that
can be expressed in terms of Hermite polynomial $H_k$:
\begin{eqnarray}
U_k(x,t) = \sqrt{\frac{{\xi }}{{\sigma_A\sqrt \pi  2^k k!}}}
(-i)^{k/2} H_k \left({\xi x/\sigma_A} \right) e^{-\xi^2x^2/2
\sigma_A^2}
 \\
V_k(y,t) = \sqrt{\frac{{\xi }}{{\sigma_F\sqrt \pi  2^k k!}}}
(-i)^{k/2} H_k \left({\xi y/\sigma_F}
 \right) e^{-\xi^2y^2/2 \sigma_F^2}.
\end{eqnarray}
In writing Eq. (\ref{Schdecomp}), we have defined:
\begin{eqnarray}
&& \mu  =  \frac{{  \sqrt {1 + (\sigma_A \sigma_F \tau)^2 }-1
}}{\tau \sigma_A \sigma_F},\\ && \xi = {\sqrt{2} {\left[ {1+
(\sigma_A \sigma_F \tau)^2 } \right]^{1/4}
 }} ,
\end{eqnarray}
with the dimensionless time $\tau  \equiv g t /2$.

With the help of Eq. (\ref{Schdecomp}), we treat $m \to x$ and $n
\to y$ and obtain an approximate Schmidt decomposition of the
state Eq. (6). Specifically, we have the Schmidt eigenvalues
\begin{equation}
\lambda_k \approx (1 - \mu^2) \mu^{2k}
\end{equation}
which decreases exponentially with mode index $k$, and the Schmidt
eigenvectors,
\begin{eqnarray}
&&  \left| {u _k } \right\rangle _a  \approx {\sum\limits_{m = -
j}^j {U_k \left( {m-m_0,t} \right)\left| \tilde m \right\rangle _a
} } ,\\ && \left| {v _k } \right\rangle _f  \approx
{\sum\limits_{n = - \infty
 }^\infty  {V_k \left( {n-n_0,t} \right)\left|
 \tilde n \right\rangle _f }},
\end{eqnarray}
where $| \tilde m \rangle _a \equiv e^{in_0(m_0-2m)\tau}  | m
\rangle_a$ and $| \tilde n \rangle _f \equiv e^{im_0(n_0-2n)\tau}  |
n \rangle_f$ are defined. The orthogonality relations between
Schmidt eigenvectors can be satisfied approximately:
\begin{eqnarray}
&& _a\langle u_k| u_l \rangle_a = \sum\limits_{m=-j}^{j} {U_k^* U_l
}\approx \int\limits_{}^{} {} dm U_k^* U_l = \delta_{kl} \\ &&
_f\langle v_k| v_l \rangle_f = \sum\limits_{n=-\infty}^{\infty}
{V_k^* V_l }\approx \int\limits_{}^{} {} dn V_k^* V_l = \delta_{kl},
\end{eqnarray}
provided that $U_k$ and $V_k$ change smoothly with integer
arguments. Therefore Eq. (19) and (20) are limited to modes with low
$k$'s, which is sufficient at short times when the expansion is
dominated by low modes.

It is interesting that the widths of the gaussian factor in $U_k$
and $V_k$ decrease with time, due to the time-dependent scale
factor $\xi$. Such a narrowing is not obvious because the left
side of Eq. (11) seems to have widths that are constant in time.
However, by taking the Fourier transform of the $e^{-x^2 /
\sigma_A^2-y^2 / \sigma_F^2} e^{ - i g x y t}$ with respect to
$x$, one can see that the transformed function is `squeezed' in a
variable consisting a linear combination of $\tilde x$ and $y$,
where $\tilde x$ is the Fourier variable of $x$. The same feature
can also been seen if one takes the Fourier transform with respect
to $y$. As $x$ ($y$) corresponds to atom (photon) number, its
Fourier variable may be interpreted as a phase degree of freedom
conjugate to the number variable. The narrowing found here
therefore reflect a time-dependent number-phase correlation
developed between two subsystems \cite{eberly}.

The entanglement entropy can now be calculated in closed form:
\begin{equation}\label{S}
{\cal S}(t) \approx  - \left[ {\frac{{\mu ^2 }}{{1 - \mu ^2 }}\ln
\left(
 {\mu ^2 } \right) + \ln \left( {1 - \mu ^2 } \right)} \right]
\end{equation}
with the time-dependence of $\mu$ given by Eq. (14). The effective
Schmidt number has a simpler expression:
\begin{equation}\label{K}
{\cal K}(t) \approx \sqrt {1 +  (\sigma_A \sigma_F \tau)^2 }
\end{equation}
which is an increasing function of the dimensionless time $\tau$.

Eq. (\ref{S}) and  (\ref{K}) describe how the degree of entanglement
increases with time. In Fig. 2, we illustrate the time dependence of
the entanglement entropy as given by Eq. (\ref{S}) and compare it
with the exact results obtained from numerical Schmidt
decomposition. We found a very good agreement in the time interval
from zero up to a certain break time $\tau_B \approx
1/\sigma_{max}$, where  $ \sigma_{max} ={{\max \{ \sigma _A , \sigma
_F \} }}$. The existence of a break time is understood from the fact
that $U_k$ ($V_k$) changes more rapidly with $m$ ($n$) as time
increases, and hence our continuous variable approximation becomes
invalid at long times. Indeed, at the break time $\tau=\tau_B$, the
width of the gaussian factor $U_k$ or $V_k$ is of the order 1, which
is minimum separation between integers.

\begin{figure}
\includegraphics[width=7 cm]{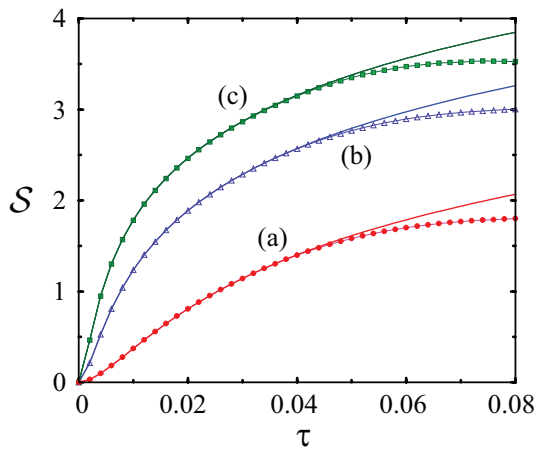}
\caption{(Color online) An illustration of the time-dependence of
entanglement entropy. The solid lines show the results from  Eq.
(\ref{S}), and the lines with data points are obtained by exact
numerical Schmidt decomposition. The initial atomic state is an
atomic spin coherent state with $N_A=2\sigma_A^2$ atoms, and the
fields are in coherent states. The following parameters are used:
(a) $\sigma_F=24$, $\sigma_A=3$, $n_0=0$ and $m_0$=0. (b)
$\sigma_F=24$, $\sigma_A=10$, $n_0=12$ and $m_0$=2. (c)
$\sigma_F=24$, $\sigma_A=18$, $n_0=0$ and $m_0$=0.  All the curves
are with the same value of $\max\{\sigma_A,\sigma_F\}=24$ so that
break time $\tau_B\approx0.04 $ is the same. } \label{fig2}
\end{figure}

\begin{figure}
\includegraphics[width=7 cm]{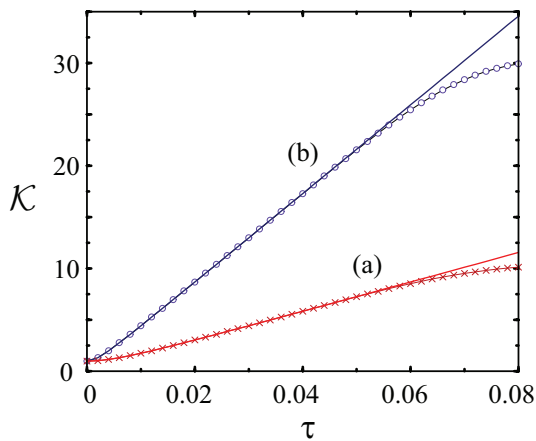}
\caption{(Color online) An illustration of time-dependence of
effective Schmidt number. The solid lines show the results from Eq.
(\ref{K}), and the line with data points are obtained by exact
numerical Schmidt decomposition. The parameters are: (a)
$\sigma_F=24$, $\sigma_A=6$, $n_0=0$ and $m_0$=0. (b) $\sigma_F=24$,
$\sigma_A=18$, $n_0=0$ and $m_0$=0.} \label{fig3 }
\end{figure}

In Fig. 3, we illustrate the time-dependence of the effective
Schmidt number ${\cal K}(t)$ as given by Eq. (\ref{K}). We see that
${\cal K}(t)$ grows almost linearly with time with the rate
$\sigma_A \sigma_F$. The results also shows a good agreement with
exact numerical calculation before the break time. We point out that
at $\tau=\tau_B$,  ${\cal K}$ reaches $\sqrt {1 + \sigma_{min}^2} $
( where $ \sigma_{min} ={{\min \{ \sigma _A , \sigma _F \} }}$),
which is independent of the peak positions $n_0$ and $m_0$.

Having discussed the entanglement in an ideal cavity, let us turn
to systems with a leaky cavity. By injecting photons into the
cavity, the question is how the output photons are entangled with
atoms. According to the Hamiltonian (2), we notice that the
interaction effectively shift the resonant frequencies of the
cavity. For a given atom number difference $m$, which corresponds
to the eigenvalue of $J_z$, the frequency shift is $g(N_A /2 \pm
m)$ for the $\pm$ polarizations. We can generalize the
input-output relations for field operators \cite{walls} to account
for the shifts, which reads,
\begin{equation}
a_{ \pm , {\rm out}} (\omega ;m) = e^{i 2\theta _ \pm  (\omega
;m)} a_{ \pm ,{\rm in}} (\omega ;m)
\end{equation}
where
\begin{equation}
e^{i2\theta _ \pm  (\omega ;m)} = \frac{{\kappa_c  + i\delta \mp ig
 m}}{{  \kappa_c  - i\delta  \pm i gm}}.
\end{equation}
Here we have defined $\delta=\omega-\omega_c-gN_A/2$, with
$\omega_c$ being the resonant frequency of the cavity, and the
cavity field decay rate $\kappa_c$. Note that the detuning $\delta$
has included the term $gN_A/2$.

In the bad cavity limit $\kappa_c \gg gm$, we find that $\theta _
\pm (\omega ;m)$ takes a simple form:
\begin{equation}
\theta _ \pm (\omega ;m)  \approx \theta _0 (\omega) \mp
\frac{gm}{\kappa_c}.
\end{equation}
provided that $\kappa_c^2 \gg \delta^2$. Here $ \theta _0
\left(\omega\right) = \tan^{-1} ( \delta / \kappa_c )$ is a phase
angle independent of $m$, and the second term $gm/ \kappa_c$ is
independent of $\omega$. Eq. (25) indicates that if the input field
frequency $\omega$ is near $\omega_c+gN_A/2$ compared with the
cavity width, then the output field operator would pick up the phase
angle $ \mp gm/ \kappa_c$, in addition to $\theta_0 (\omega)$. Such
a result can be translated into Schrodinger picture. Suppose the
input field is described by a single mode in the Fock state $|N_+,
N_- \rangle_{{\rm in}}$ and the atomic states is the state $|m
\rangle_a$, then we have the transformation for the input-output
state vectors \cite{singlemode}:
\begin{equation}
|N_+, N_- \rangle_{{\rm in}} |m \rangle_a \to e^{i2 \theta_0
(\omega) s -i2gmn/\kappa_c} |N_+, N_- \rangle_{{\rm out}} |m
\rangle_a
\end{equation}
where $n=N_+-N_-$ and $s=N_++N_-$. Apart from an unimportant factor
$e^{i2 \theta_0 (\omega) s}$ that is not responsible for quantum
entanglement, the correlated phase factor $e^{-i2gmn/\kappa_c}$ is
precisely of the same form as Eq. (6) derived in an ideal cavity.
Therefore our Schmidt analysis in the ideal cavity can be employed
by taking the interaction time $t = 2/\kappa_c$, which is the finite
life time of cavity fields. In particular, if $\tau_B \ge 2 /
\kappa_c$, we have
\begin{equation}
{\cal K} \approx \sqrt {1 + (2\sigma_A \sigma_F g/\kappa_c)^2 }
\end{equation}
after all the input photons are fully converted into the output
mode.

It is interesting to employ Eq. (27) to estimate the entanglement
generated in free space, i.e., without the cavity. This may be done
by replacing $g$ with the corresponding value in free space $g_f$,
and $2/\kappa_c$ by the (single pass) interaction time $t_f$.
Apparently, $g > g_f$ due to the cavity enhancement would increase
the entanglement for a fixed interaction time.  A crude analysis by
Kuzmich {\it et al.} in Ref. \cite{sqz1} suggested that $g_f$ is
inversely proportional to the volume defined by the spatial extend
of the atoms in free space. As $g$ is inversely proportional to the
cavity mode volume, $g/g_f$ could be approximated by the volume
ratio. However, a rigorous analysis in free space is difficult
because of the involvement of multi-mode dynamics in general
\cite{sqz1}.

To conclude, we perform the Schmidt analysis of the photon-atom
entanglement generated by Faraday rotation in a cavity. We present
an approximate analytic formula of Schmidt decomposition, which
reveals the Schmidt mode structures and the time-dependence of
entanglement. In particular we show that the initial fluctuations
$\sigma_F$ and $\sigma_A$ are key parameters to control the rate of
change of entanglement. This work is restricted to the class of
gaussian amplitudes defined by Eq. (9) and Eq. (10), which covers a
range of quantum states of photons and atoms that have a well
defined peak and width. In this paper we also discuss the
entanglement in leaky systems when cavity decay rate is sufficiently
large. The quantum dynamics beyond the bad cavity limit is out of
the scope of this paper. This problem is difficult because the phase
shift of output photons becomes a nonlinear function of $m$ when
$\kappa_c$ is comparable with $g$. The nonlinearity, which comes
from the competition between natural cavity response function and
atom-field interaction, cannot be treated by the approximation
method described in this paper. This is an open topic for future
investigations.

\begin{acknowledgments}
This work is supported in part by the Research Grants Council of
the Hong Kong Special Administrative Region, China (Project No.
401305).
\end{acknowledgments}


\begin{references}
\bibitem{coherent} For a two-mode coherent field state
$\left| {\alpha _+ } \right\rangle \left| {\alpha _- }
\right\rangle$, $$P_{s,n}= e^{ - \langle {s/2 }\rangle }
{\frac{{\left( {\alpha_+ } \right) ^{s/2 + n/2} \left( {\alpha_- }
\right) ^{s/2 - n/2} }}{\sqrt{{\left( {s/2 + n/2}
 \right)!\left( {s/2 - n/2} \right)!}}}} $$
where $\langle s \rangle =|\alpha_+|^2+|\alpha_-|^2$ is the
average total photon number. When $\left\langle {s } \right
\rangle \gg 1$, $F _n$ defined after Eq. (5) can be well
approximated by a gaussian with $\sigma_F=2 \sqrt{\left\langle {s
} \right\rangle}$ and $n_0=|\alpha_+|^2-|\alpha_-|^2 $.
\bibitem{singlemode} A single mode field considered here
does not necessarily mean that it is purely monochromatic. In
general a (wave packet) mode corresponds to a general superposition
of frequency modes, so that the input Fock state $|N_{\pm}
\rangle_{{\rm in}} =b_ {\pm,{\rm in}}^{\dag N}|0\rangle / \sqrt
{N_{\pm}!}$ is defined by the operator: $b_{ \pm ,{\rm in}} =
\int\limits_{}^{} {f(\omega )a_{ \pm ,{\rm in}} (\omega )d\omega }$.
A similar definition can be written for the output mode. For the
state transformation (26) to be valid, $f(\omega)$ has to
concentrated at $\omega_c+gN_A/2$ and its width is small compared
with $\kappa_c$.
\end{references}
\end{document}